\documentclass[%
 reprint, superscriptaddress,
 amsmath,amssymb,
 aps,
showkeys,
]{revtex4-2}

\usepackage{graphicx}
\usepackage{booktabs}
\usepackage{longtable}
\usepackage{siunitx}
\usepackage[version=4]{mhchem}
\usepackage[colorlinks=true,linkcolor=blue,citecolor=blue,urlcolor=blue]{hyperref}
\usepackage{multirow}

\newcommand{\isotope}[2]{$^{\text{#2}}$#1}

\mathchardef\ordinarycolon\mathcode`\:
\mathcode`\:=\string"8000
\begingroup \catcode`\:=\active
  \gdef:{\mathrel{\mathop\ordinarycolon}}
\endgroup

\usepackage{lineno}

\begin{document}

\title{Zone refining of ultra-high purity sodium iodide for low-background detectors}

\author{Burkhant Suerfu}
\email{suerfu@alumni.princeton.edu}
\affiliation{University of California, Berkeley, Department of Physics, Berkeley, CA 94720, USA}

\author{Frank Calaprice}
\affiliation{Department of Physics, Princeton University, Princeton, NJ, 08544, USA}

\author{Michael Souza}
\affiliation{Department of Physics, Princeton University, Princeton, NJ, 08544, USA}
\affiliation{Department of Chemistry, Princeton University, Princeton, NJ, 08544, USA}

\date{\today}

\begin{abstract}

There has been a growing interest in ultra-high purity, low-background NaI(Tl) crystals for dark matter direct searches. Past research indicates that zone refining is an efficient and scalable way to purify NaI. In particular, K and Rb---two elements with radioisotopes that can cause scintillation backgrounds---can be efficiently removed by zone refining. However, zone refining has never been demonstrated for ultra-high purity NaI which became commercially available recently. In this article, we show that many common metallic impurities can be efficiently removed via zone refining. A numerical model for predicting the final impurity distribution was developed and used to fit the ICP-MS measurement data to determine the segregation coefficient and the initial concentration. Under this scheme, the segregation coefficient for K is estimated to be $0.57$, indicating that zone refining is still effective in removing K from ultra-high purity NaI. As zone refining tends to move the impurities to one end, elements with concentrations too low to be measured directly in the unprocessed powder can potentially be detected in the end due to the enrichment. We also present an analysis technique to estimate the initial concentrations of impurities with partial data, which effectively enhances the sensitivity of the spectrometer. Using this technique, the initial concentration of \isotope{Rb}{85} is estimated to be between 5~ppt and 14~ppt at 90\% CL, at least 14~times lower than the detection limit of ICP-MS and 7~times lower than the current most stringent limit set by the DAMA collaboration by direct counting of radioactive \isotope{Rb}{87}. These results imply that zone refining is a key technique in developing next-generation, NaI-based crystal scintillators for dark matter direct detection.


\end{abstract}

\maketitle


\section{Introduction} \label{sec:intro}

Thallium-activated sodium iodide NaI(Tl) crystal is one of the most important and widely used scintillators in the fields of radiation detection and gamma spectroscopy. In recent years, there has been a growing interest in ultra-high purity NaI(Tl) crystals for dark matter searches~\cite{dama-libra,anais,cosine100,sabre-pop,cosinus}. In these experiments, trace amounts of radioactive impurities---in particular $^{40}$K, $^{210}$Pb and $^{3}$H---can lead to background events in the energy region of interest. Ultra-high purity NaI powder with sub-ppm metallic impurities is commercially available, but the purity level is not yet sufficient for efficient dark matter searches and further purification is desired.

Since the pioneering work of Pfann in 1952~\cite{pfann1952}, zone refining has been used to purify silicon and germanium for the semiconductor industry. Later, this technology was extended to other elements and compounds, including alkali halide salts. In 1960, Gross showed that many alkali and alkali earth impurities can be separated and removed from alkali iodides by zone refining~\cite{gross}. However, due to the limitations of the assay technology, his work was done with dopants at 100~ppm. The effectiveness of zone refining at the ppb level remains to be demonstrated. In this study, we investigate the effectiveness of zone refining in purifying commercial ultra-high purity NaI powder and develop relevant quantitative simulation and data analysis techniques to estimate the segregation coefficients~($k$) and the average initial impurity concentrations~($C_0$) of common impurities in the NaI powder.

\section{Experimental Procedure}

For this study, \SI{744}{\gram} of low-potassium, ultra-high purity NaI powder from Sigma-Aldrich was carefully and thoroughly dried and sealed in a carefully-cleaned synthetic fused silica crucible and zone-refined for 53~passes over approximately two weeks. 

\subsection{Crucible Preparation and Powder Drying}
To avoid contamination by impurities diffusing out of the crucible~\cite{verhoogen1952ionic,suerfu-thesis}, a 1.5-inch-diameter, 4-ft-long synthetic fused silica crucible was used to seal and contain the ingot during the zone refining~\cite{suerfu-thesis}. The crucible was cleaned by rinsing with 5\% by volume hydrofluoric acid followed by 5\% by volume high-purity hydrochloric acid and deionized water. Subsequently, the crucible was vacuum-baked at \SI{400}{\celsius} to remove surface-adsorbed water.

To remove residual water from the ultra-high purity powder, the powder was first loaded into a 3-inch-diameter, 2-ft-long fused silica tube in a glove box purged with boil-off nitrogen and vacuum-baked for one week at several stages of increasing temperatures starting at room temperature. A detailed operational procedure is outlined in~\cite{suerfu-thesis}.

\subsection{Powder Drying and Treatment with Silicon Tetrachloride}

\begin{figure}[htb!]
    \centering
    \includegraphics[width=\linewidth]{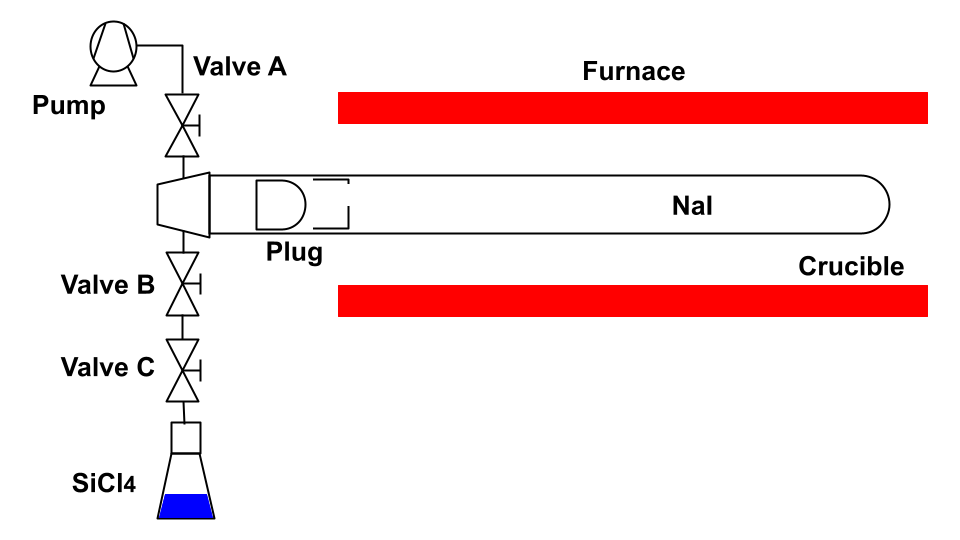}
    \caption{The \ce{NaI} powder was melted under \ce{SiCl4} atmosphere prior to zone refining. The plug on the right was used to introduce \ce{SiCl4} vapor through the vent hole on the topside while preventing the backflow of the molten NaI. The plug on the left was used to hermetically seal the crucible once the powder has been treated with \ce{SiCl4}.}
    \label{fig:setup}
\end{figure}

Trace amounts of \ce{Na2O} and \ce{NaOH} can cause the ingot to stick to the crucible and make the extraction difficult. To remove \ce{Na2O} and \ce{NaOH} already present in the powder prior to drying, the dried NaI powder was further melted under silicon tetrachloride~(\ce{SiCl4}) atmosphere in the cleaned and dried crucible to be used for zone refining~\cite{suerfu-thesis,eckstein}. The setup used is illustrated in Fig.~\ref{fig:setup}.

The NaI powder was first loaded into the crucible inside the glove box, and two fused silica plugs---one with a vent hole---were placed inside the crucible. A Pyrex vacuum adapter with two valves was attached to the open end of the crucible by compression fitting before taking out of the glove box. The assembly was placed horizontally and pumped to high vacuum using a turbo-molecular pump via one of the two valves on the Pyrex adapter~(Valve~A). Once the vacuum level reached $10^{-5}$~mbar, the plug with the vent hole was fused to the crucible with a torch. A 50-mL flask containing \ce{SiCl4} was attached to the other valve on the adapter~(Valve~B) and the section containing the NaI powder was inserted into a horizontal furnace. The temperature of the furnace was slowly increased to \SI{800}{\celsius}. 
When the temperature reached \SI{450}{\celsius}, Valve~A is closed and the valves between the \ce{SiCl4} and the NaI powder~(Valve~B and Valve~C) were opened to introduce the \ce{SiCl4} vapor. After 1~hour at \SI{800}{\celsius}, the oven temperature was ramped down to \SI{400}{\celsius} and the crucible was sealed using the other fused silica plug so that some \ce{SiCl4} vapor remained in the crucible during zone refining.

\subsection{Zone Refining}
\begin{figure}[htb!]
    \centering
    \includegraphics[width=\linewidth]{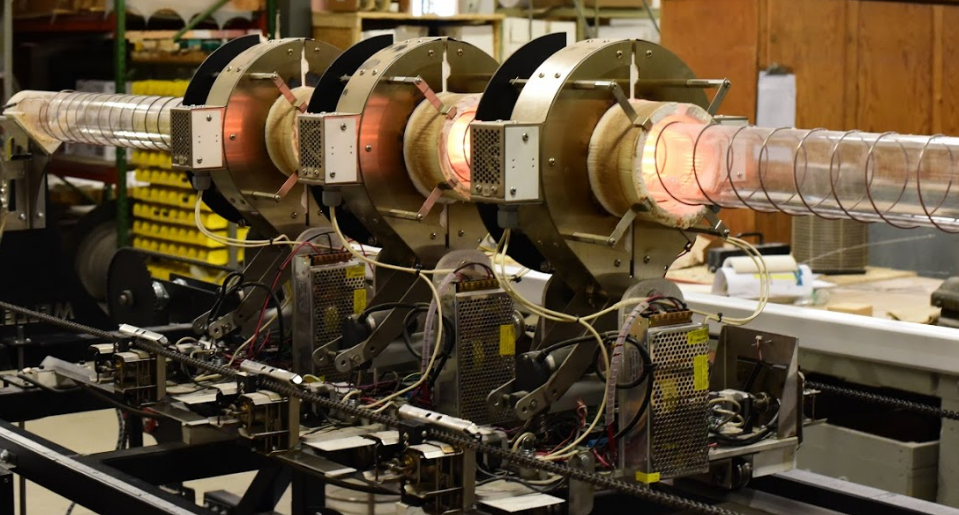}
    \caption{Zone refining in progress at Mellen Company with three zones. The crucible is placed inside a fused silica preheater tube maintained at \SI{300}{\celsius} by a resistive wire wound around the preheater tube.}
    \label{fig:zone refining at mellen}
\end{figure}

The sealed crucible was taken to Mellen Company in Concord, New Hampshire for zone refining. In total, 53 zone passes were carried out at a rate of 2~in./hr~(Fig.~\ref{fig:zone refining at mellen}). After zone refining, the crucible was brought back to Princeton. Samples were cut from different locations of the ingot and sent to Seastar Chemicals for assay with inductively-coupled plasma mass spectroscopy~(ICP-MS). Prior to measurement, the samples were first rinsed with high-purity ethanol and deionized water to remove surface contaminants introduced during the sample cutting, packing and shipment. The distribution of impurity concentration was then compared to the expected distribution from numerical model to estimate the segregation coefficient~$k$ and the average initial concentration~$C_0$.

\section{Numerical Model}\label{sec:model}

\begin{figure}[htb!]
    \centering
    \includegraphics[width=\linewidth]{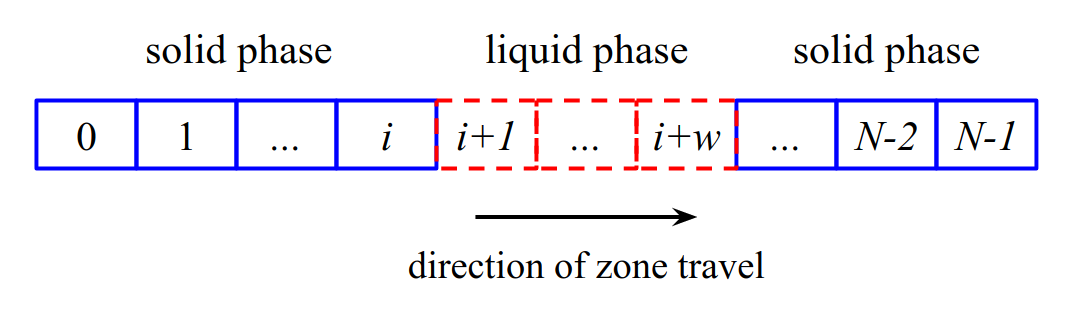}
    \caption{The ingot is divided into $N$ cells with equal width. Solid blue cells represent solidified sections of the ingot and the dashed red cells represent the molten liquid zone. Two arrays, $A_j$ and $C_j$ are used to denote the cross section and the solute concentration of the $j$-th cell. When the zone's left boundary travels from $i$ to $i+1$, cell~$i$ solidifies and its solute concentration is given by $k$ times the solute concentration of the molten liquid zone~(Eq.~\ref{eq:conc}). As the right boundary advances to ($i+w$)-th cell, the molten zone's solute concentration and total volume is updated based on the volume and the amount of solutes in the newly melted cell~(Eq.~\ref{eq:solute} and Eq.~\ref{eq:volume}). To simulate the entire zone refining process, the molten zone is swept across the entire ingot incrementally and the appropriate boundary relation is applied at each step.}
    \label{fig:algorithm}
\end{figure}

Since there is no established analytical solution of the solute distribution for an arbitrary cross section profile and an arbitrary number of zone passes, we developed a numerical model to simulate the final solute distribution~\cite{suerfu-thesis}. In this numerical model, the entire ingot is divided into multiple cells with equal width, as illustrated in Fig.~\ref{fig:algorithm}. To make the simulation easier, the overall length of the ingot is normalized to the total number of cells~$N$, and the width of the molten zone~$w$ is expressed as the integer number of cells spanned by the zone~(see Fig.~\ref{fig:algorithm}).

Let $A_j$ and $C_j$ denote the cross section and the solute concentration of the $j$-th cell. Two variables, $M$ and $V$, are used to keep track of the total amount of solute and the total volume of the molten zone. Assuming cells~$i+1$ to $i+w$ are molten as in Fig.~\ref{fig:algorithm}, the following relations exist:
\begin{eqnarray}
    C_i &\leftarrow& k\frac{M}{V}, \label{eq:conc}\\
    M &\leftarrow& M - A_iC_i + A_{i+w}C_{i+w}, \label{eq:solute}\\
    V &\leftarrow& V - A_i + A_{i+w}, \label{eq:volume}
\end{eqnarray}
where $\leftarrow$ is the assignment operator, and $k$ is the (effective) segregation coefficient, defined as the ratio of solute concentrations in the solid and the liquid phase. Note that since the width of each cell is identical, it is set to unity and has been left out. Equation~\ref{eq:conc} sets the solute concentration in the solidified $i$-th cell based on the definition of $k$, whereas Eq.~\ref{eq:solute} and Eq.~\ref{eq:volume} update $M$ and $V$ with the loss of solute due to the solidified $i$-th cell and the gain of solute due to the newly melted ($i+w$)-th cell. To simulate one cycle of zone purification, the above relations are applied incrementally to each cell starting from the beginning. Special care is taken at both ends: at the beginning of the ingot, $M$ and $V$ are initialized to the sum of the respective quantities in the first $w$ cells since $M$ and $V$ will keep rising until the first zone solidifies:
\begin{eqnarray}
    M &\leftarrow& \sum_{i=0}^{w-1}A_iC_i,\\
    V &\leftarrow& \sum_{i=0}^{w-1}A_i.
\end{eqnarray}

At the other end, the terms representing the gain due to the newly melted cell in Eq.~\ref{eq:solute} and Eq.~\ref{eq:volume} are left out since no new material is melted as the zone moves:
\begin{eqnarray}
    M &\leftarrow& M - A_iC_i, \\ 
    V &\leftarrow& V - A_i.
\end{eqnarray}

\begin{figure}[!htb]
    \centering
    \includegraphics[width=\linewidth]{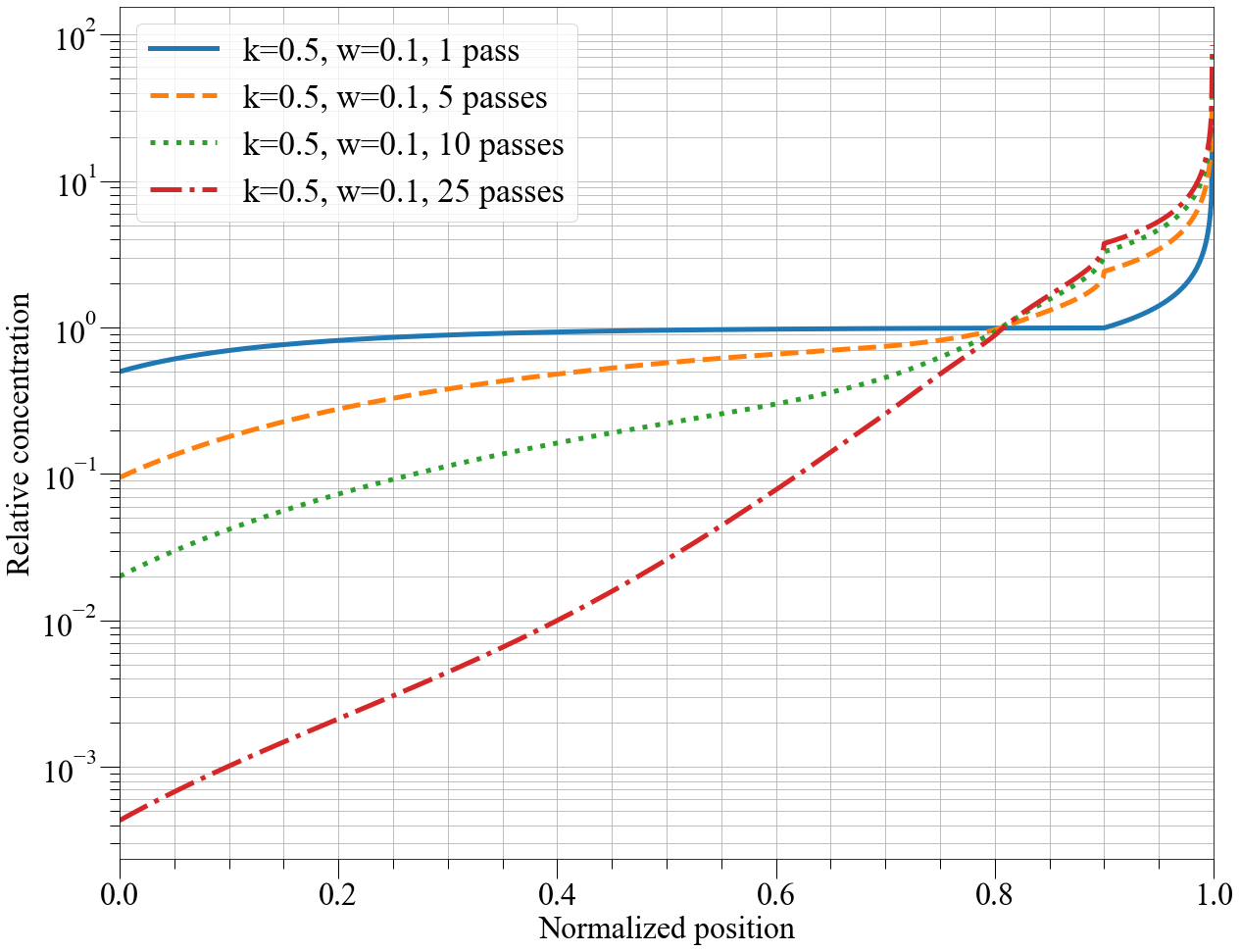}
    \caption{Numerically calculated solute distributions after 1, 5, 10 and 25 zone passes. The segregation coefficient is assumed to be $0.5$ and zone width 10\% of the ingot length. The initial average solute concentration is normalized to 1. As zone refining progresses, more and more impurities are pushed to the end of the ingot.}
    \label{fig:zr demo}
\end{figure}

The above operation can be applied repeatedly to simulate multiple zone passes. As an illustration, Fig.~\ref{fig:zr demo} shows the solute distributions in an ingot with uniform cross section after different numbers of zone passes. With this numerical approach, it is also straightforward to account for non-uniform initial solute distribution and non-uniform ingot cross section~(e.g. the tip of a crystal grown by the vertical Bridgman method where the crucible is tapered). This algorithm is a reasonable description of the zone refining process when 1) the segregation coefficient is constant over the solute concentrations of interest, 2) solute redistribution is dominated by zone refining instead of vapor transport or diffusion, and 3) change of cross section profile in the process is negligible.

\section{Data Analysis}

Since the unprocessed powder had impurity concentrations already near the detection threshold, the concentrations of many elements in the first half of the ingot were expected to be below the ICP-MS detection limit. In addition, the concentrations of most solutes were expected to vary more rapidly towards the end of the ingot. Therefore, instead of taking samples from equally-spaced positions, one sample was taken from the beginning of the ingot and other four samples were taken from roughly equally-spaced positions in the latter half of the ingot. In the numerical model, the zone width was assumed to be 2~inches based on visual observation of the zone refining process, and the number of zone passes was taken to be 53, consistent with the experiment. The cross section profile of the ingot after zone refining was measured carefully and used in the numerical model. 

\begin{figure}[htb!]
    \centering
    \includegraphics[width=\linewidth]{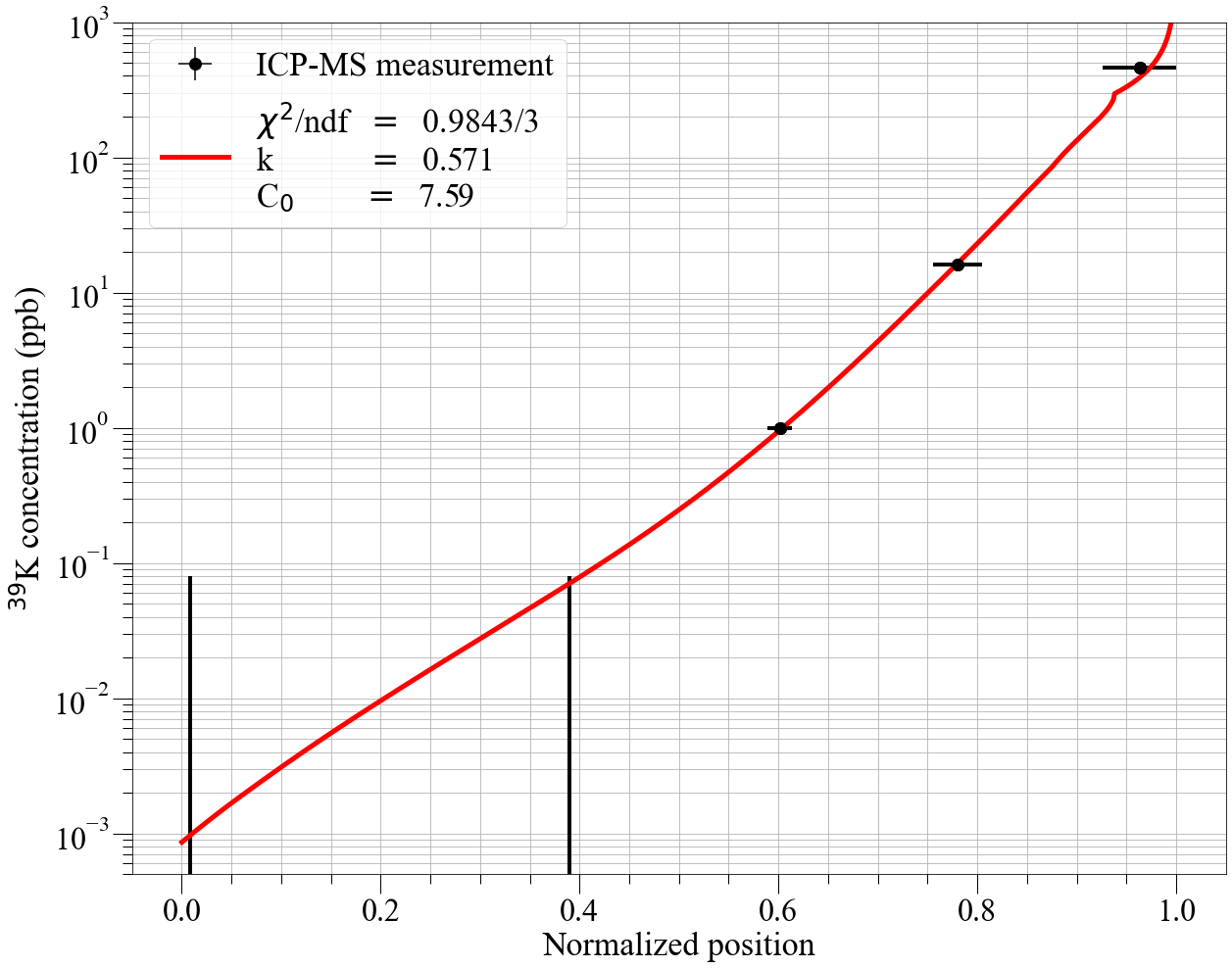}
    \caption{Distribution of the concentration of \isotope{K}{39} as a function of the position. The position is measured from the beginning of the ingot where zone refining started. The geometrical center of each sample is taken as the x~coordinate of the data point and the size of the sample as the x~error. The solid curve represents the best fit to the data.}
    \label{fig:k-distrib}
\end{figure}

To estimate $k$ and $C_0$ of various elements in NaI, the numerical model was used to fit the measured position-dependent impurity concentration of each element by minimizing $\chi^2$. Good to excellent agreements were obtained between the data and the best-fit model for many elements with a few exceptions. Figure~\ref{fig:k-distrib} shows as an example the ICP-MS data and the best-fit model distribution of \isotope{K}{39}.

\begin{figure}[htb!]
    \centering
    \includegraphics[width=\linewidth]{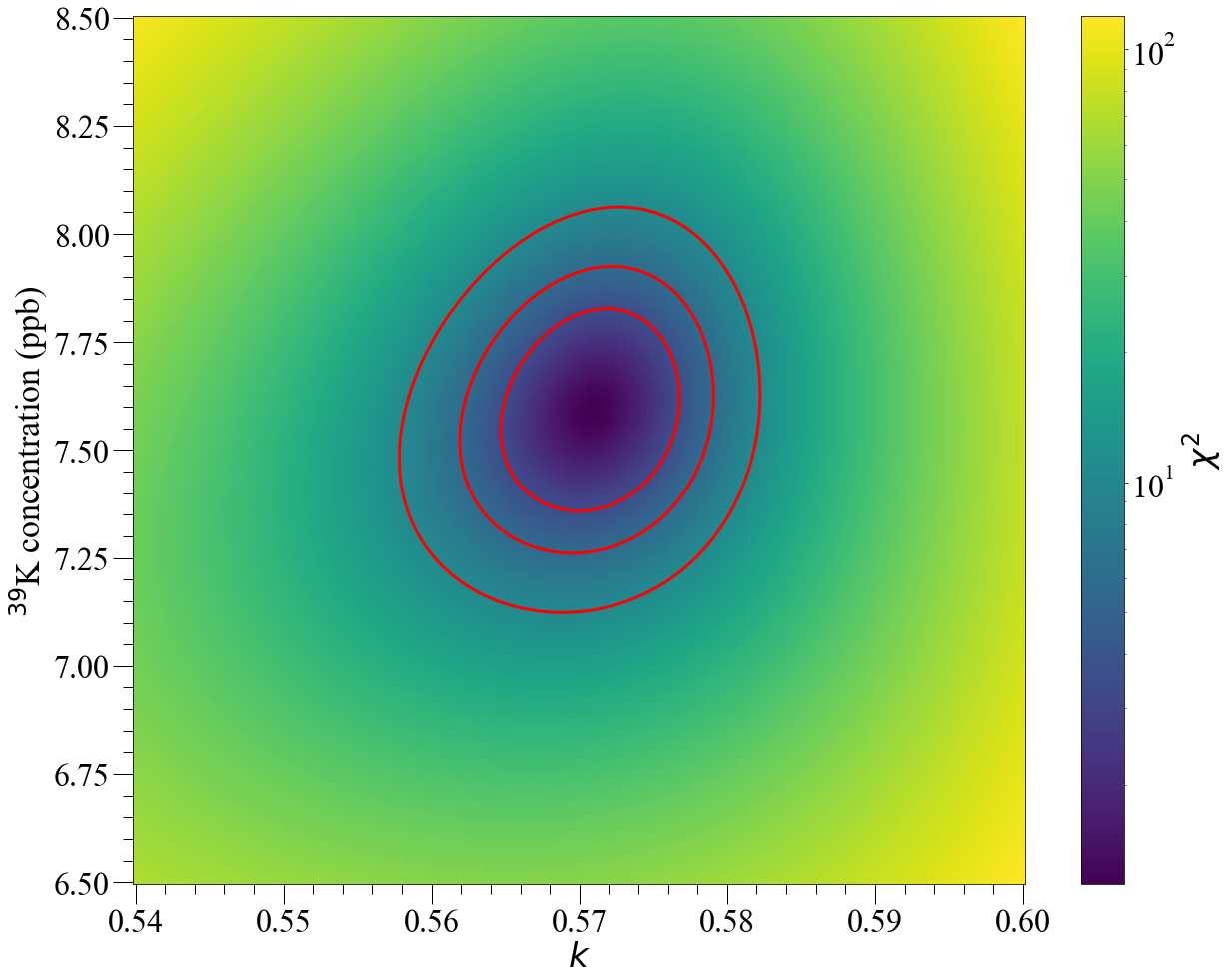}
    \caption{\isotope{K}{39} $\chi^2$ map obtained by grid search in the $k$-$C_0$ space. The $\chi^2$ map is characterized by a well-defined global minimum, since \isotope{K}{39} concentration is above the spectrometer detection limit at three different locations, allowing for simultaneous determination of $k$ and $C_0$. The three contour lines correspond to 1-$\sigma$, 90\% and 99\% confidence levels.}
    \label{fig:chi2map k39}
\end{figure}

To avoid local minima and to get an accurate estimate of the uncertainties, the parameters corresponding to the minimum $\chi^2$ were determined by a grid search in the $k$-$C_0$ space. Once the global minima was located, the boundary of the $\Delta\chi^2=2.3$ contour was used as the associated upper and lower 1-$\sigma$ errors. The $\chi^2$ map of \isotope{K}{39} obtained as the result of grid search is illustrated in Fig.~\ref{fig:chi2map k39} along with 1-$\sigma$, 90\%~CL and 99\%~CL contours.

\begin{figure}[htb!]
    \centering
    \includegraphics[width=\linewidth]{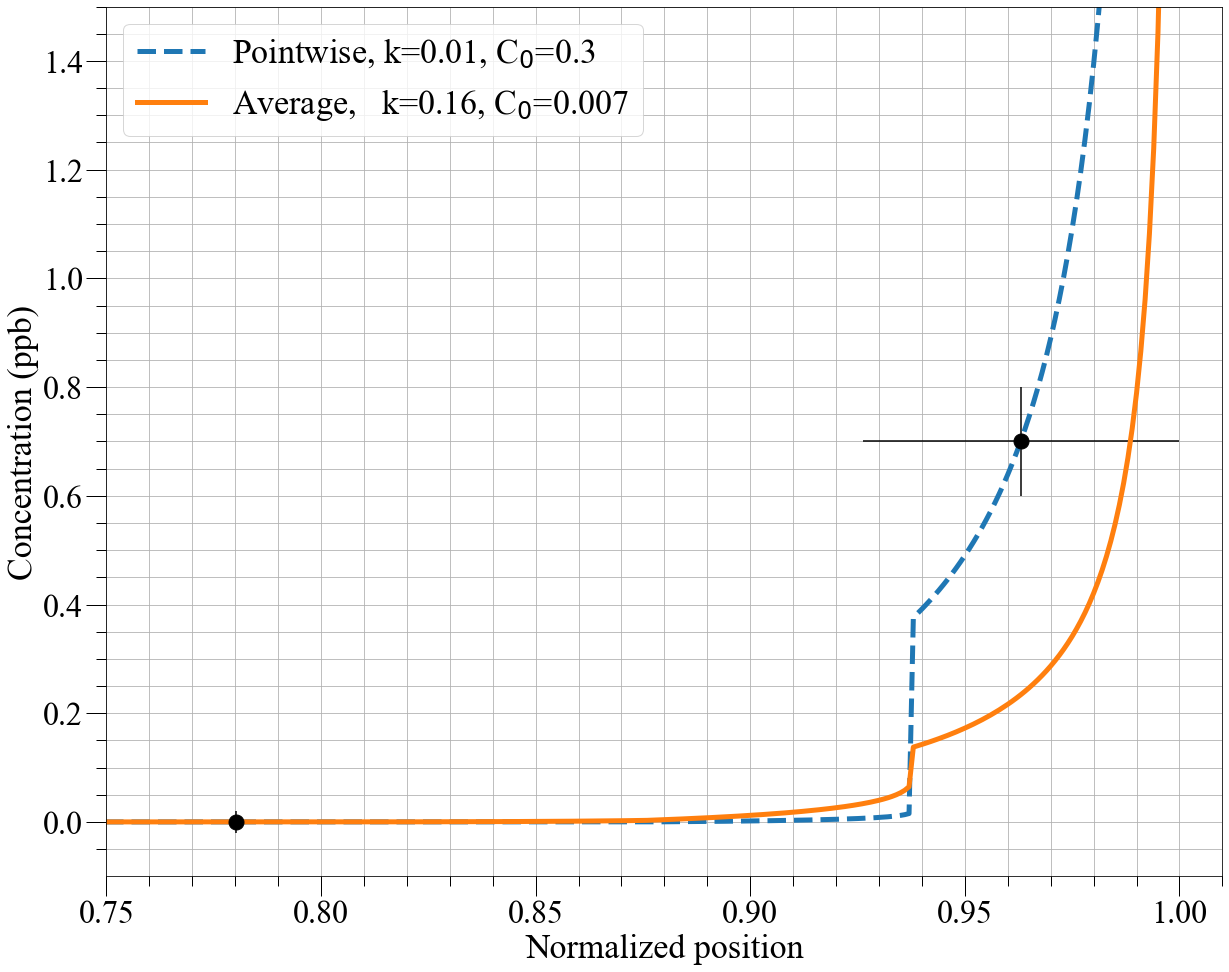}
    \caption{Pointwise comparison of \isotope{Rb}{85} data leads to overestimation of $C_0$. \isotope{Rb}{85} was detected only in the last sample at the end of the ingot after 53~zone passes. Since the concentration varies rapidly near the end, the true concentration-weighed center differs significantly from the geometrical middle of the sample. When the middle of the last sample is used as the x~coordinate in a pointwise $\chi^2$ computation, the distribution that matches exactly at the last data point is favored. Since the first four data points are all upper limits, $k$ can take an arbitrarily small value at the cost of increased $C_0$, thereby leading to overestimation of $C_0$~(dashed blue line). To properly treat this case, $\chi^2$ should be computed using the average of the model distribution~(solid orange line).}
    \label{fig:rb-distrib}
\end{figure}

When computing the $\chi^2$, the effect of each sample's finite geometrical size was taken into account by fitting the average of the model distribution over the physical dimension of the sample instead of pointwise comparison. When there were more than two data points, the two approaches yielded similar results. However when there was only one data point and the other four data points were upper limits, pointwise comparison would lead to the divergence of $C_0$. This is illustrated in Fig.~\ref{fig:rb-distrib} using \isotope{Rb}{85} data.

\subsection{Treatment of Partial Data}\label{sec: analysis upper limit}

When there are more than two data points with definite values, both parameters can be determined with relative ease: the relative difference between the two points sets $k$ and the absolute scale sets $C_0$. Figure~\ref{fig:chi2map k39} shows the $\chi^2$ map for the \isotope{K}{39} distribution in Fig.~\ref{fig:k-distrib} where three data points are available. The $\chi^2$ map is characterized by a well-defined global minimum.

When there is only one data point at the end of the ingot and all other data points are upper limits, it is not possible to determine either parameters from the fit. This is because the available information cannot be used to determine the ``inefficiency'' of the zone refining: two values of $k$ cannot be distinguished as long as they are both sufficiently small to move enough amount of impurity such that the concentrations at other sample locations are below the detection limit.

\begin{figure}[htb!]
    \centering
    \includegraphics[width=\linewidth]{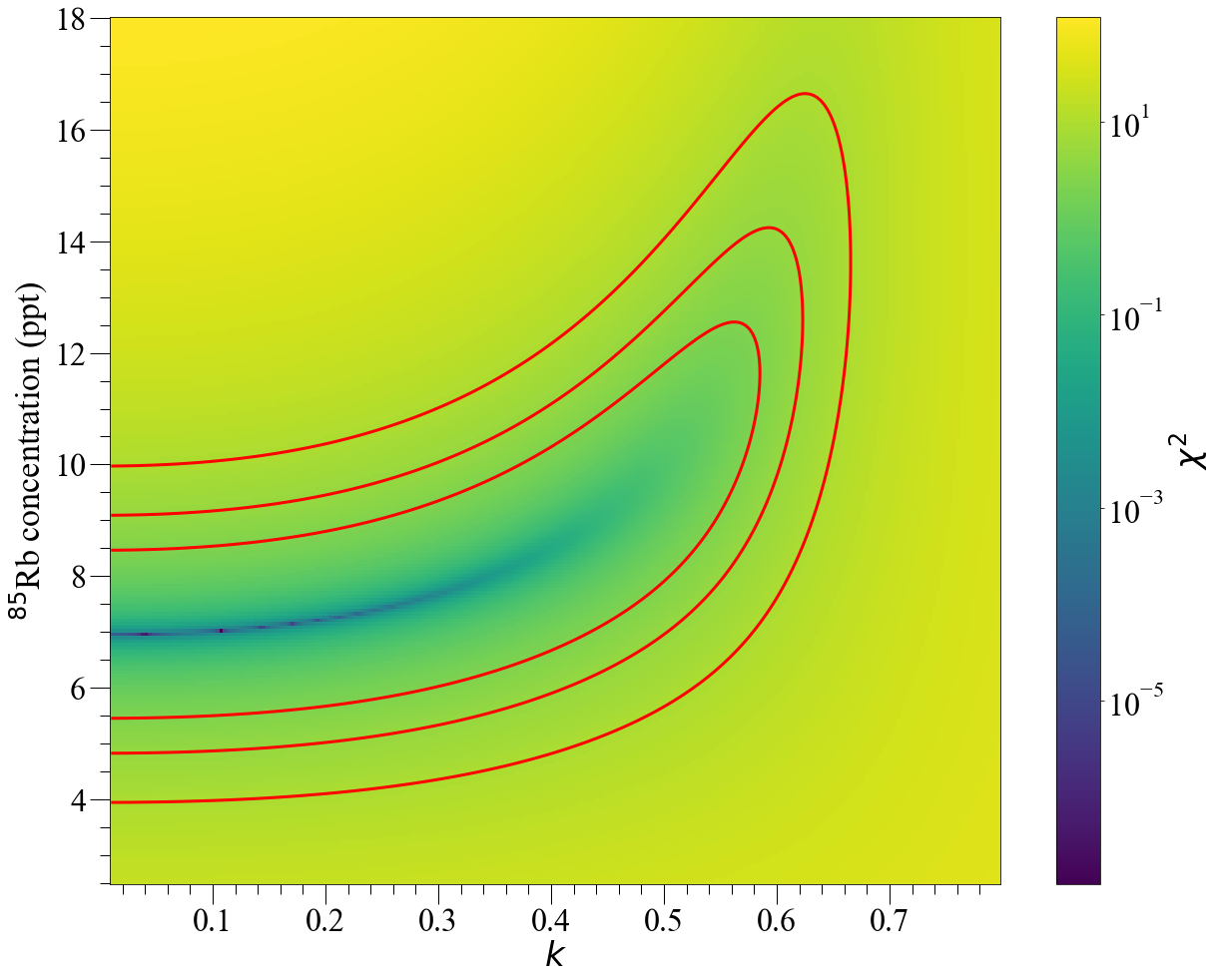}
    \caption{The $\chi^2$ map for \isotope{Rb}{85} is characterized by a band and a series of local minima since there is only one data point. If there is one more data point, $k$ will be constrained from below by the inefficiency of the zone refining and the $\chi^2$ map will resemble Fig.~\ref{fig:chi2map k39}. The three contour lines correspond to 1-$\sigma$, 90\% and 99\% confidence levels.}
    \label{fig:chi2map rb85}
\end{figure}

However, even in this scenario an upper limit for $k$ and a confidence interval for $C_0$ can be found. This is because $C_0$ is constrained from below---the total amount of impurity cannot be less than that detected in the last sample alone. If the upper limits are small compared to the last data point,  zone refining is very efficient and the majority of the impurities should be concentrated on the end. Thus, the amount of impurity remaining in other parts of the ingot are several orders of magnitude smaller and the associated errors are small as well. In other words, the uncertainty in $C_0$ due to the uncertainty in $k$ is small. This is illustrated by the flat band of $C_0$ as $k$ goes to 0 in Fig.~\ref{fig:chi2map rb85}. In this case, a relatively narrow confidence interval for $C_0$ can be derived.

\section{Results}

Fifty six metal and semimetal elements were analyzed using ICP-MS. Although only a few isotopes are relevant in the context of NaI as a low-background scintillator, for completeness the best-fit parameters, estimated confidence intervals or detection limits are summarized in Table~\ref{tab:det} and Table~\ref{tab:undet} in the Appendix. 

Of the isotopes assayed, K, Rb and Pb are of particular interest for NaI-based low-background detectors since these elements have radioactive isotopes that can cause scintillation background events by the emissions of low-energy electrons and/or X-rays~\cite{anais-background,cosine-background}. The measurement results for these isotopes are summarized in Table~\ref{tab: isotopes of interest measurement} and the analysis results are summarized in Table~\ref{tab: isotopes of interest}.

\begin{table}[h!]
    \centering
    \caption{The measured concentrations of impurities of interest at the five sample locations. The range after the $\pm$ sign corresponds to the physical size of each sample. A few other isotopes are included to show that zone refining also works for other elements as well.}
    \label{tab: isotopes of interest measurement}
    \newcommand{\ra}[1]{\renewcommand{\arraystretch}{#1}}
    \ra{1.3}
    \begin{tabular}{ccccccc}
        \toprule
        \toprule
        \multirow{3}{*}{\textbf{Isotope}} &
        \multicolumn{6}{c}{\textbf{Impurity concentration~(ppb)}} \\ \cline{2-7}
        & \multirow{2}{*}{Powder}& \multicolumn{5}{c}{Sample location~(mm)}\\
        & & 7$\pm$7 & 325$\pm$9 & 492$\pm$10 & 635$\pm$20 & 783$\pm$30\\
        \cline{1-7}

        \isotope{K}{39} & 7.5 & $<$0.8 & $<$0.8 & 1 & 16 & 460\\
        \isotope{Rb}{85} & $<$0.2 & $<$0.2 & $<$0.2 & $<$0.2 & $<$0.2 & 0.7\\
        \isotope{Pb}{208} & 1.0 & 0.4 & 0.4 & $<$0.4 & 0.5 & 0.5\\
        \isotope{Cu}{65} & 7 & $<$2 & $<$2 & $<$2 & 2 & 620\\
        \isotope{Cs}{133} & 44 & 0.3 & 0.2 & 0.5 & 23.3 & 760\\
        \isotope{Ba}{138} & 9 & 0.1 & 0.2 & 1.4 & 19 & 330\\
        \bottomrule
        \bottomrule
    \end{tabular}
\end{table}

\begin{table}[h!]
    \centering
    \caption{The average initial concentrations and segregation coefficients of the isotopes of interest.}
    \label{tab: isotopes of interest}
    \newcommand{\ra}[1]{\renewcommand{\arraystretch}{#1}}
    \ra{1.3}
    \begin{tabular}{*4c}
        \toprule
        \toprule
        \textbf{Isotope} & \textbf{$k$} & \textbf{$C_0$ (ppb)} & \textbf{Powder~(ppb)}\\
        \midrule

        \isotope{K}{39} & 0.571$\pm$0.006 & 7.6$\pm$0.2 & 7.5\\
        \isotope{Rb}{85} & $<$0.59 & (0.005, 0.014) & $<$0.2\\
        \isotope{Pb}{208} & \multicolumn{2}{c}{no observable separation} & 1.0\\
        \bottomrule
        \bottomrule
    \end{tabular}
\end{table}


Potassium has a primordial radioisotope, \isotope{K}{40} that can decay by the capture of a K-shell electron to \isotope{Ar}{40} with 10.7\% branching ratio. When the accompanying high-energy $\gamma$ ray escapes the crystal without interaction, the crystal sees 3-keV X-rays/auger electrons, which is in the middle of the region where the dark matter annual modulation signal is reported~\cite{dama-libra}. Due to its large natural abundance, ubiquitous presence and similar chemical properties to Na, \isotope{K}{40} is a source of background for \isotope{K}{nat} concentration as low as $\sim$10~ppb.

In this study, K was shown to be effectively separated from NaI by zone refining. Figure~\ref{fig:k-distrib} shows the concentration of \isotope{K}{39} as a function of position for the zone-refined ingot. It can be seen that most K was pushed to the end of the ingot. The segregation coefficient was estimated to be 0.57, which is higher than that reported by Gross~\cite{gross} but is in good agreement with other independent measurements by single crystal growth~\cite{suerfu-thesis,suerfu-nai033}. The agreement with $k$ obtained via crystal growth indicates that at a zone traveling rate of 2~in./hr, the measured effective segregation coefficient is already very close to the true segregation coefficient. $C_0$ estimated by the fit is also in excellent agreement with the measurement value of the unprocessed powder.


The primordial radionuclide of rubidium---\isotope{Rb}{87}---is a pure $\beta$ emitter with an end point energy of 282~keV. Although it has a long half life~($4.8\times 10^{10}$ yr), a high natural abundance of 27.8\% makes it a potential background for NaI-based dark matter detectors even at trace levels~\cite{suerfu-thesis}. ICP-MS measurement of Rb is often performed on the stable isotope \isotope{Rb}{85}. However, typical sensitivity to \isotope{Rb}{85} is around 0.2~ppb while \isotope{Rb}{87} can still contribute a significant background at this level.

In this work, the sensitivity to \isotope{Rb}{85} was further enhanced by the enrichment of \isotope{Rb}{85} in the last sample via zone refining. Using the analysis outlined in Sec.~\ref{sec: analysis upper limit}, $k$ was constrained to be below 0.58 and $C_0$ between 5 to 14~ppt at 90\%~CL~(Table~\ref{tab:det} and Fig.~\ref{fig:chi2map rb85}).


The radioisotope of lead---\isotope{Pb}{210}---is one of the biggest backgrounds to low-background NaI(Tl) detectors~\cite{cosine-background,anais-background,suerfu-nai033}. Unlike \isotope{K}{40} and \isotope{Rb}{87}, \isotope{Pb}{210} is produced by the decay of \isotope{Rn}{222} and its concentration is not correlated to other stable isotopes of lead. Nonetheless, how stable \isotope{Pb}{208} redistributes during zone refining can indicate the effectiveness of removing \isotope{Pb}{210}. In the 5 samples studied, the concentrations of \isotope{Pb}{208} showed a uniform reduction compared to that of the powder, but the distribution did not follow the characteristic distribution of zone refining, suggesting that either zone refining is not very effective in removing lead, or there is an alternative, more important transport mechanism for the redistribution of lead. This will be investigated further in a separate study.

\section{Discussions}

This study shows that zone refining can be used to purify ultra-high purity NaI powder to further remove K and Rb impurities very effectively. In addition, the process of zone refining can be employed as a means to enhance the sensitivity of ICP-MS to measure impurities with concentrations normally below the detection limit of the spectrometer.

\subsection{Improving Crystal Purity}

\begin{figure}[htb!]
    \centering
    \includegraphics[width=\linewidth]{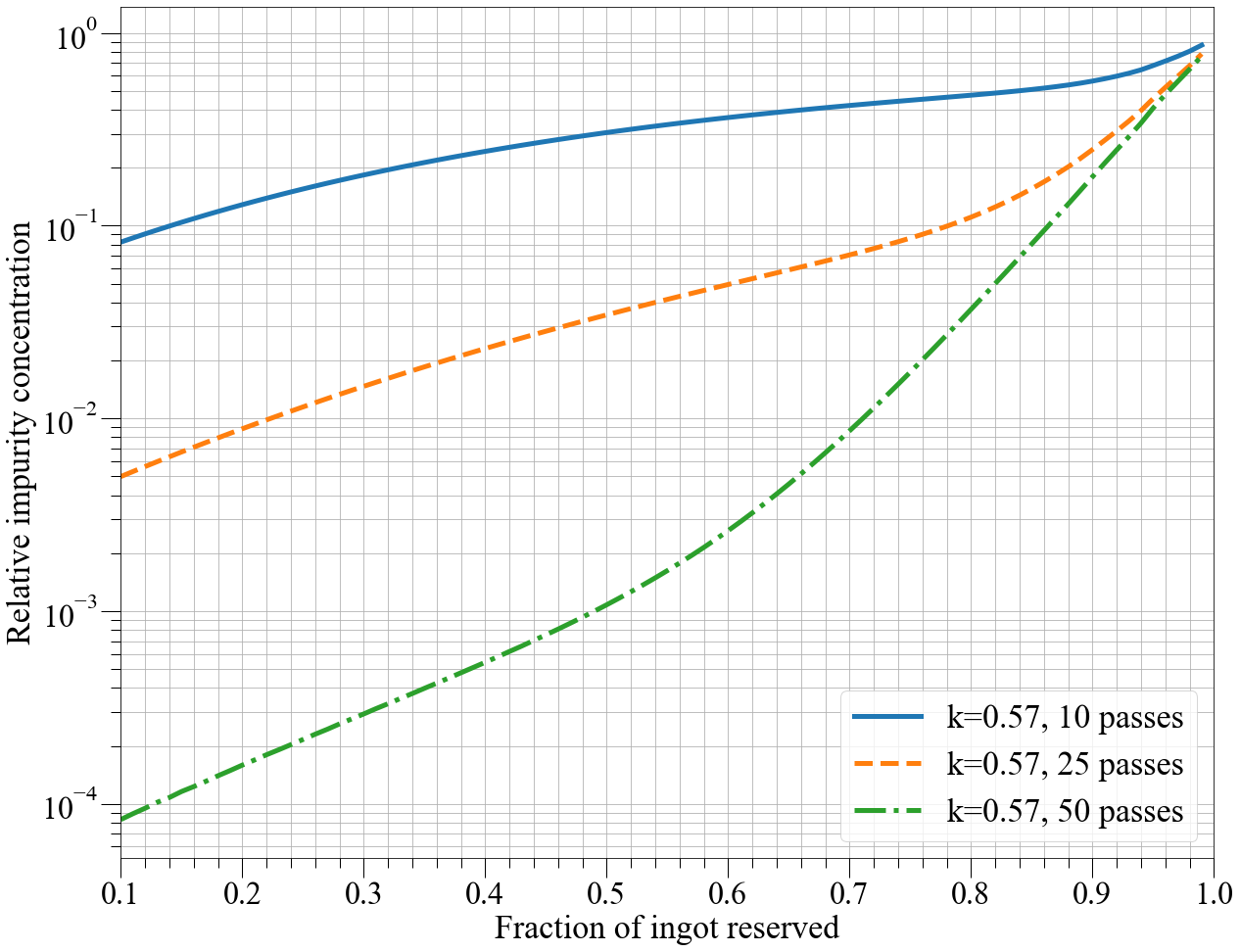}
    \caption{Impurity reduction as a function of the fraction of material kept after zone refining after 10, 25, and 50 zone passes. The zone width is assumed to be 6\% of the length of the ingot, and the segregation coefficient is set to be 0.57.}
    \label{fig:pure factor}
\end{figure}

The purification power of zone refining depends on the segregation coefficient~$k$ as well as the fraction of the material kept. The smaller $k$ is, the more efficiently impurities move to the end of the ingot, and the more the dirty part of the ingot is discarded, the higher the average purity of the remaining material will be. Figure~\ref{fig:pure factor} shows the final relative impurity~(K) concentration as a function of the fraction of the ingot reserved after different numbers of zone passes. It indicates that after 50 zone passes, the purity of 70\%~(85\%) of the ingot can be improved by a factor of 100~(25). Since the upper limit of $k$ for Rb is also close to 0.57, we expect Rb to be removed at least as efficiently as K during zone refining, and much more efficiently if $k$ is on the order of 0.1 as measured by \cite{gross} and estimated from thermodynamic phase diagram~\cite{suerfu-thesis, fact-database}.

\subsection{Enhancing Measurement Sensitivity}

In the process of zone refining, the impurities are moved towards the ends of the ingot to varying degrees. Due to this enrichment effect, the impurities in the ends can be more easily detected and measured. Given a reasonable model of the system, the original concentration can be constrained using the process outlined in Sec.~\ref{sec: analysis upper limit}.

In this study, we constrained the original concentration of \isotope{Rb}{85} in the powder to be between 5 and 14~ppt, at least 14~times lower than the detection limit of ICP-MS. The initial concentration of \isotope{Rb}{87} inferred from \isotope{Rb}{85} is 1.9 to 5.4~ppt. The corresponding activity is 14.4 to \SI{40.4}{\micro\becquerel/\kg}, at least 7~times lower than the most stringent upper limit of \SI{300}{\micro\becquerel/\kg} obtained by the DAMA/LIBRA collaboration by direct counting~\cite{dama-apparatus}.

\subsection{Impact on Dark Matter Searches}

Although recently, the technology to grow large NaI(Tl) crystal with \isotope{K}{39} concentration as low as 4.3~ppb~(4.6~ppb \isotope{K}{\text{nat}}) has been demonstrated~\cite{suerfu-nai033}, most NaI(Tl)-based dark matter detectors have K concentrations around 30~ppb~\cite{anais-background,cosine-background}. The projected background due to 30~ppb of \isotope{K}{nat} in the absence of active veto is approximately 0.5~cpd/kg/keV~\cite{anais-background, cosine-background,sabre-mc}. If zone refining is used with 70\% yield, the background due to \isotope{K}{40} can be reduced to 0.005~cpd/kg/keV, which is smaller than 0.01~cpd/kg/keV, the amplitude of the annual modulation.

Unlike K, the concentration of Rb is naturally low in NaI, and no background due to \isotope{Rb}{87} has been reported yet. In this study, we constrained for the first time the concentration of \isotope{Rb}{85} in the ultra-high purity NaI powder. The inferred \isotope{Rb}{nat} concentration is about 7 to 19~ppt, and the associated \isotope{Rb}{87} background is less than 0.02~cpd/kg/keV in the NaI powder. Even without zone refining, crystal growth will further reduce this background. Therefore in practice, Rb in ultra-high purity NaI powder is not a significant source of background.

It must be pointed out that the above estimates are concerned with impurities in the NaI powder. When NaI is doped with TlI to make NaI(Tl) crystal scintillator, the impurities in TlI will typically be diluted by a factor of 1000. If zone refining is used to purify NaI powder of ultra-high purity, at one point, the final purity will be dominated by the impurities in the TlI dopant instead of NaI.

\section{Conclusions}

Zone refining was performed on ultra-high purity NaI powder sealed inside a synthetic fused silica crucible. By measuring impurity concentrations at different locations of the ingot and comparing them to the numerical model, the segregation coefficients and initial impurity concentrations were calculated for several elements. In addition, using the enrichment effect of zone refining, we also constrained for the first time the concentration of Rb in ultra-high purity NaI powder, at a level at least 14~times lower than the detection limits of ICP-MS and 7~times lower than the previous measurement by direct counting of radioactive \isotope{Rb}{87}. During the zone refining, two sources of background, K and Rb, are shown to be efficiently separated from NaI. Calculations indicate that zone refining can easily reduce K and Rb by a factor of 10 to 100. This indicates that zone refining is an effective and scalable technique to purify ultra-high purity NaI powder for next-generation low-background NaI detectors.




\section*{Acknowledgements}
The authors would like to thank Jonathan Mellen and Ashley Mellen at the Mellen Company for help operating the zone refining system, and Brad McKelvey for help with ICP-MS assays. The authors would also like to thank Aldo Ianni, Masayuki Wada, and Xuefeng Ding for careful reviews of this manuscript and useful discussions. This research is supported by the National Science Foundation under award
number PHY-1242625, PHY-1506397 and PHY-1620085.
\section*{Appendix}

\begin{table}[htb!]
    \caption{Summary of the estimated $C_0$ and $k$~(best fit values and 1-$\sigma$ errors). When there is only one measurement value, the upper limit of $k$ and the range for $C_0$ are stated at 90\%~confidence level. Elements marked with~* showed increased concentrations near the dirty end, but the distributions in the clean end deviated noticeably from that expected from zone refining. Elements marked with~$\dagger$ exhibited irregular distributions. Furthermore, with the exceptions of Ga and Ca, there were noticeable reduction of the total amount of impurities after zone refining, manifesting as either noticeably smaller $C_0$ compared to the powder, or ICP-MS values of the zone-refined ingot consistently smaller than that of the powder. One explanation is that at such low concentrations, vapor transport is comparable to or more important than zone refining for these elements.}
    \label{tab:det}
    \newcommand{\ra}[1]{\renewcommand{\arraystretch}{#1}}
    \ra{1.3}
    \begin{tabular}{*4c}
        \toprule
        \toprule
        \textbf{Isotope(s)} & \textbf{$k$} & \textbf{$C_0$ (ppb)} & \textbf{Powder~(ppb)} \\
        \midrule

        \isotope{Li}{7} & 0.51$^{+0.01}_{-0.02}$ & 5.5$^{+0.2}_{-0.3}$ & 2
        \\
        \isotope{Mg$^*$}{24} & $0.63$ & 2.7 & 14
        \\
        \isotope{K}{39} & 0.571$\pm$0.006 & 
        7.6$\pm$0.2 & 7.5 
        \\
        \isotope{Cr}{52} & 0.48$\pm$0.01 & 2.1$\pm$0.2 & $<$1
        \\
        \isotope{Fe}{56} & 0.45$\pm$0.02 & 11.8$\pm1.7$ & 9
        \\
        \isotope{Cu}{65} & 0.37$^{+0.04}_{-0.09}$ & 7.3$^{+0.4}_{-0.5}$ & 7
        \\
        \isotope{Ga$^*$}{69} & 0.78 & 1.7 & 1.5
        \\
        \isotope{Cs$^*$}{133} & 0.391 & 9.1 & 44
        \\
        \isotope{Ba$^*$}{138} & 0.61 & 6.2 & 9
        \\
        \midrule
        
        \isotope{Mn}{55} & $<0.37$ & (0.29, 0.37) & $<$0.5\\
        \isotope{Co}{59} & $<$0.95 & (0.01, 0.42) & $<$0.5\\
        \isotope{Ni}{60} & $<$0.58 & (0.39, 1.10)  & $<$10\\
        \isotope{Zn}{68} & $<$0.70 & (0.08, 0.44)  & $<$13\\
        \isotope{Rb}{85} & $<$0.59 & (0.005, 0.014) & $<$0.2\\
        \isotope{Sr}{88} & $>$1.4 & (0.08, 0.44)  & $<$0.1 \\
        \isotope{Y}{89}  & $<$0.50 & (0.0, 0.05)  & $<$0.1 \\
        \isotope{Zr}{90} & $<$0.76 & (0.01, 0.10)  & $<$2
        \\
        \isotope{La}{139}& $<$0.31 & (0.15, 0.2)  & $<$1
        \\
        \isotope{Yb}{172}& $>$1 & (0.0, 0.09)  & $<$0.1
        \\
        \midrule
        \isotope{Al$^\dagger$}{27}  & \multicolumn{2}{c}{ \multirow{4}{*}{irregular distribution} } & 3\\
        \isotope{Ca$^\dagger$}{44} & & & 130\\ \isotope{Ti$^\dagger$}{48}  & & & 2\\ \isotope{Pb$^\dagger$}{208} & & & 1\\
        \bottomrule
        \bottomrule
    \end{tabular}
\end{table}

\begin{table}[htb!]
    \caption{Summary of isotopes whose concentrations are below the detection limit in all the samples.}
    \label{tab:undet}
    \newcommand{\ra}[1]{\renewcommand{\arraystretch}{#1}}
    \ra{1.3}
    \begin{tabular}{lc}
        \toprule
        \toprule
        
        \textbf{Isotope} & \textbf{$C_0$ (ppb)}\\
        \midrule
        \isotope{U}{238} & $<$0.1\\
        \isotope{Eu}{153}, \isotope{Ho}{165}, \isotope{Lu}{175}, \isotope{Au}{197} & $<$0.2\\
        \isotope{V}{51}, \isotope{Sm}{149}, \isotope{Tm}{169}, & \multirow{2}{*}{$<$0.5} \\
        \isotope{Hf}{178}, \isotope{Re}{187} & 
        \\
        \isotope{Th}{232} & $<$0.8\\
        \isotope{Be}{9}, \isotope{Nb}{93}, \isotope{Cd}{111}, \isotope{Nd}{146}, & \multirow{2}{*}{$<$1} \\
        \isotope{Gd}{157}, \isotope{Ta}{181}, \isotope{Pt}{195} & \\
        \isotope{Mo}{98}, \isotope{Ag}{107} & $<$5\\
        \isotope{B}{11}, \isotope{Ge}{72}, \isotope{As}{75}, \isotope{Pd}{105}, & \multirow{2}{*}{$<$10} \\
        \isotope{Sb}{121}, \isotope{Pr}{141}, \isotope{Er}{166} & \\
        \isotope{Bi}{209} & $<$20\\
        \isotope{Ru}{102}, \isotope{W}{182} & $<$50\\
        \isotope{Sc}{45}, \isotope{Rh}{103} & $<$100\\
        
        \bottomrule
        \bottomrule
    \end{tabular}
\end{table}

\newpage
\bibliographystyle{apsrev4-1}


\bibliography{reference.bib}

\end{document}